\documentclass[%
 reprint,
superscriptaddress,
nofootinbib,
 amsmath,amssymb,
prl
longbibliography
]{revtex4-2}

\usepackage{makecell} % for table
\usepackage[colorinlistoftodos]{todonotes} % for todo
\usepackage{natbib}

\usepackage{subfiles}
\usepackage{float}
\usepackage{graphicx}% Include figure files
\usepackage{dcolumn}% Align table columns on decimal point
\usepackage{bm}% bold math
\usepackage{multirow}
\usepackage{appendix}
\usepackage{lipsum} % TODO: commend it out later
\usepackage{amsmath}
\let\oldAA\AA
\renewcommand{\AA}{\text{\normalfont\oldAA}}

\begin{document}

% \preprint{APS/123-Quantum Chemistry}

\title{Electric Polarization from Many-Body Neural Network Ansatz}
\author{Xiang Li}
\email{lixiang.62770689@bytedance.com}
\affiliation{ByteDance Research, Zhonghang Plaza, No. 43,  North 3rd Ring West Road, Haidian District, Beijing.}
\author{Yubing Qian}
\email{X. L. and Y. Q. contributed equally to this work.}
\affiliation{ByteDance Research, Zhonghang Plaza, No. 43,  North 3rd Ring West Road, Haidian District, Beijing.}
\affiliation{%
School of Physics, Peking University, Beijing 100871, People’s Republic of China
}
\author{Ji Chen}
\email{ji.chen@pku.edu.cn}
\affiliation{%
School of Physics, Peking University, Beijing 100871, People’s Republic of China
}
\affiliation{%
Interdisciplinary Institute of Light-Element Quantum Materials, Frontiers
Science Center for Nano-Optoelectronics, Peking University, Beijing 100871, People’s Republic of China
}

\date{\today}% It is always \today, today,
\begin{abstract}
    \textit{Ab initio} calculation of dielectric response with high-accuracy electronic structure methods is a long-standing problem, for which mean-field approaches are widely used and electron correlations are mostly treated via approximated functionals.
    %which remains less investigated for decades. Despite the importance of the electron correlation, the mean-field approach is widely preferred in polarization calculations to save computational costs. 
    Here we employ a neural network wavefunction ansatz combined with quantum Monte Carlo to incorporate correlations into polarization calculations. 
    %Our approach is validated on a variety of systems, 
    On a variety of systems, including isolated atoms, one-dimensional chains, two-dimensional slabs, and three-dimensional cubes,
    the calculated results outperform conventional density functional theory and are consistent with the most accurate calculations and experimental data. 
    %Furthermore, we have utilized our method to evaluate the out-of-plane dielectric constant of bilayer graphene and investigate its thickness dependence.
    Furthermore, we have studied the out-of-plane dielectric constant of bilayer graphene using our method and re-established its thickness dependence.
    Overall, this approach provides a powerful tool to consider electron correlation in the modern theory of polarization.
\end{abstract}

\maketitle

\begin{figure}[b]
\centering
\includegraphics[width=0.5\textwidth]{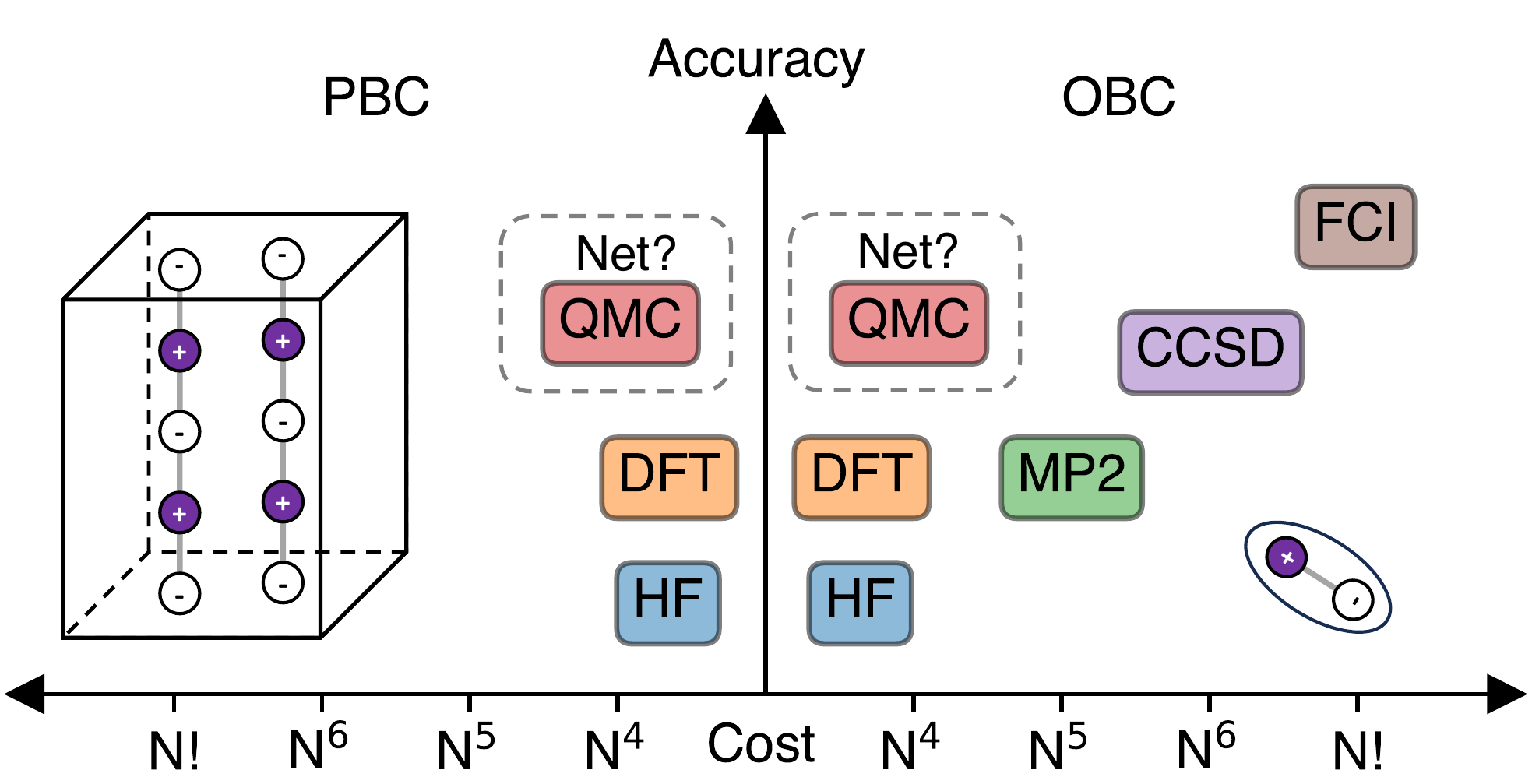}
\caption{\label{fig:concept} A brief illustration of the computational cost and accuracy of different electronic structure methods in polarization calculations. $N$ denotes the number of electrons in the system. High-level correlated wavefunction methods are not shown in the PBC panel because they have not been applied to PBC polarization calculations so far. QMC: quantum Monte Carlo; DFT: density functional theory; HF: Hartree-Fock; MP2: second-order M{\o}ller-Plesset perturbation theory; CCSD: coupled cluster with single and double excitations; FCI: full configuration interaction.}
\end{figure}

Electric polarization plays a crucial role in electromagnetic phenomena such as ferroelectricity and piezoelectricity. 
Despite its significance, a proper microscopic definition of polarization was only formulated in the 1990s  
\cite{king-smith_theory_1993,resta_macroscopic_1994}, which revealed the hidden relation between physical polarization and the Berry phase of solid systems. 
This theoretical advance leads to successful calculations of the dielectric response of solid materials from first principles \cite{nunes_berry-phase_2001,souza_first-principles_2002,chain_slab_dft}, which is critical in several fields of condensed matter physics, such as the ferroelectric and topological materials \cite{xiao_berry_2010}.
However, the underlying electronic structure methods are mostly mean-field approaches such as density functional theory (DFT) \cite{kohn_nobel_1999}, which has its limitation because the result depends heavily on the so-called exchange-correlation functional. 
Exchange-correlation functionals can not fully account for the exact correlation effects of electrons.
In particular, widely used semi-local functionals often produce an excessive overestimate of electric susceptibility \cite{alpha_qc,chain_slab_dft}.
Although correlated wavefunction methods, such as coupled-cluster theory 
%with single, double, and perturbative triple excitations [CCSD(T)], 
can also be employed to calculate polarization \cite{chain_qc}, their high computational complexity hinders their application in solid systems.
Furthermore, most of these correlated electronic structure methods are limited in open boundary conditions (OBC) for polarization calculations, which leads to slow convergence and heavy computational costs towards the thermodynamic limit (TDL), see Fig.~\ref{fig:concept} for a summary of the state-of-the-art methods in polarization calculations.

In addition to the conventional deterministic electronic structure methods mentioned above, quantum Monte Carlo (QMC) methods are also widely adopted for electronic structure calculations, showing favorable computational scaling and high accuracy \cite{qmc_heg,qmc_rev,fciqmc_solid}. 
Pioneering works to study electric susceptibility 
using QMC have been reported \cite{qmc_polar,umari_linear_2009}, in which traditional Slater Jastrow type wavefunction is combined with diffusion Monte Carlo (DMC) to study polarization of hydrogen chains in periodic boundary conditions (PBC). 
The main difficulty for DMC is to write down the local self-consistent Hamiltonian under a finite electric field and run calculations iteratively. 
Despite the promising results on hydrogen chain \cite{qmc_polar},
there are still grand challenges: multiple loops of DMC are needed for the self-consistent procedure, a complex forward walking strategy is required for evaluating the polarization, and the quality of trial wavefunction affects the accuracy of DMC. 
Therefore, it is desirable to develop more accurate and efficient approaches to calculate the electric polarization of solid systems.

In recent years, there has been significant progress in the application of neural network in the electronic structure community. Neural network wavefunction ansatz combined with QMC simulations has demonstrated higher accuracy with lower computational complexity than conventional high-order wavefunction methods~\cite{carleo_science,nomura_restricted_2017,carleo_constructing_2018,deepwf,ferminet,paulinet,rbm_mol,excited,deeperwin,nqs_rev,wapnet,ferminet_heg,boson_periodic,deepsolid,rbm_solid,force_net,ferminet_ecp,ferminet_dmc,xie2022deep}.
The expressiveness of neural network overcomes the main bottleneck of traditional wavefunction ansatz in QMC, 
making the approach a competitive option for state-of-the-art electronic structure calculation.
So far, the neural network QMC calculations have shown great power in treating spin systems \cite{carleo_science,nomura_restricted_2017,carleo_constructing_2018}, molecules \cite{deepwf,ferminet,paulinet,rbm_mol}, periodic models \cite{wapnet,ferminet_heg,boson_periodic,deepsolid}, and real solids \cite{deepsolid,rbm_solid}.

In this work, we extend the neural network QMC calculation to the electric polarization of solid systems.
Specifically, we employ a recently developed solid neural network, dubbed DeepSolid \cite{deepsolid}, in conjunction with variational Monte Carlo (VMC).
Antithetic sampling \cite{monte_carlo_book} is developed for efficient computation of the Berry phase and thus the electric polarization.
Our approach has been tested on a diverse range of systems, including isolated atoms, one-dimensional chains, two-dimensional slabs, three-dimensional cubes, and bilayer graphene. The results demonstrate clear advantages of our approach over traditional methods.

\begin{table*}[htb]
\caption{\label{tab:atom} Calculated atom polarizability in atomic unit\ $({\rm Bohr}^3)$. DeepSolid results are labeled as DS. B3LYP, HF, and CCSD(T) results are calculated with PySCF \cite{pyscf} in the def2-qzvppd basis set and non-relativistic limit. Recommended data is taken from Ref.~\citep{atom_polar}, which is deduced from experiment data and the most accurate calculations.}
% \begin{ruledtabular}
% \begin{tabular}{ccccccccccc|c} 
%     %\hline
%       & H & He & Li & Be & B & C & N & O & F & Ne & MAE\\
%      \hline
%     B3LYP & 5.187 & 1.485 & 142.727 & 43.090 & 26.923 & 13.408 & 7.711& 5.180 & 3.563 & 2.838 & 3.684 \\
%     HF & 4.484 & 1.318 & 169.231 & 45.441& 25.541 & 12.462 & 7.138 & 5.180 & 3.082 & 2.365 & 2.044 \\
%     CCSD(T) & 4.484 & 1.372 & 165.803 & 37.707 & 24.174 & 12.293 & 7.212 & 5.787 & 3.347 & 2.642 & 0.751 \\
%     DS & 4.511(4) & 1.400(1) & 165.0(1) & 37.58(3) & 18.90(2) & 12.11(2) & 7.189(8) & 5.236(8) & 3.776(7) & 2.705(4) & 0.384 \\
%     Recommended & 4.5(exact) & 1.38375(2) & 164.1125(5) & 37.74(3) & 20.5(1) & 11.3(2) & 7.4(2) & 5.3(2) & 3.74(8) & 2.66110(3) & 0 \\
% \end{tabular}
% \end{ruledtabular}
% \end{table*}
\begin{ruledtabular}
\begin{tabular}{ccccccc|c} 
    %\hline
      & H & He & Li & Be & N & Ne & MAE\\
     \hline
    B3LYP & 5.187 & 1.485 & 142.727 & 43.090 & 7.711 & 2.838 & 4.669 \\
    HF & 4.484 & 1.318 & 169.231 & 45.441 & 7.138 & 2.365 & 2.243 \\
    CCSD(T) & 4.484 & 1.372 & 165.803 & $\mathbf{37.707}$ & $\mathbf{7.212}$ & 2.642 & 0.326 \\
    DS & $\mathbf{4.51}$ & $\mathbf{1.39}$ & $\mathbf{165.0}$ & 36.95 & 7.16 & $\mathbf{2.67}$ & 0.32 \\
    Recommended & 4.5(exact) & 1.38375(2) & 164.1125(5) & 37.74(3) &  7.4(2) & 2.66110(3) & 0 \\
\end{tabular}
\end{ruledtabular}
\end{table*}

 To introduce our methodology, let us consider a crystal system under a finite electric field $\mathbf{E}$, the enthalpy of this system is formulated below \cite{qmc_polar,nunes_real-space_1994,nunes_berry-phase_2001,souza_first-principles_2002}
\begin{equation}
\begin{gathered}
    F[\psi]=\frac{\langle\psi|\hat{H}_S|\psi\rangle}{\langle\psi|\psi\rangle}-\Omega_S\mathbf{E}\cdot \mathbf{P}[\psi]\ ,
\end{gathered}
\label{eq:loss}
\end{equation}
where $\hat{H}_S$ denotes the supercell Hamiltonian in the absence of electric field $\mathbf{E}$, and $\Omega_S$ is the supercell volume. The term $-\Omega_S\mathbf{E}\cdot\mathbf{P}$ represents the interaction between electric polarization density $\mathbf{P}$ and electric field. However, a proper microscopic definition of $\mathbf{P}[\psi]$ remains absent for decades since the ordinary position operator $\hat{\mathbf{r}}$ violates the periodic boundary condition. This problem was finally solved after recognizing the polarization as the Berry phase in the Brillouin zone, according to which the polarization can be extracted from a general wavefunction $\psi$ as follows \cite{resta_quantum-mechanical_1998}
% However, the calculation of $\mathbf{P}$ is not straightforward because the ordinary position operator $\hat{\mathbf{r}}$ violates the periodic boundary condition and does not yield properly defined macroscopic polarization. This obstacle was not removed until MTP was established, according to which the polarization $\mathbf{P}$ should be calculated from the Berry phase of the corresponding ansatz $\psi$. 
% To calculate the Berry phase, we utilize the single-point formula proposed by Resta \cite{resta_quantum-mechanical_1998} to avoid integrating over the $k$-space. 
% For a general correlated wavefunction:
%
\begin{equation}
\begin{gathered}
    \mathbf{P}[\psi]=-\frac{1}{\Omega_S}\sum_{i}\frac{\mathbf{a}_i}{2\pi}{\rm Im}\ln\frac{\langle\psi|\hat{U}_i|\psi\rangle}{\langle\psi|\psi\rangle}\ ,\\
    \hat{U}_i=\exp\left[\mathbf{i}\mathbf{b}_i\cdot\left(\sum_e\hat{\mathbf{r}}_e-\sum_I Z_I\mathbf{R}_I\right)\right]\ ,
\end{gathered}
\label{eq:phase}
\end{equation}
where $\mathbf{a}_i,\mathbf{b}_i$ denote lattice and reciprocal lattice vectors of the supercell. $\hat{U}_i$ serves as a periodic generalization of the position operator $\hat{\mathbf{r}}$ in solid systems and ${\rm Im \ln}(x)$ is used to extract the Berry phase within $x$. Note that $\hat{U}_i$ is an intrinsic many-body operator which includes all the electron coordinates in the exponent. A charge-weighted sum of ion coordinates $Z_I\mathbf{R}_I$ is also included to achieve translation invariance of polarization.

With the enthalpy functional formulated above, traditional methods usually start with a Hartree-Fock (HF) ansatz, which is typically expressed as follows
\begin{equation}
\begin{gathered}
    \psi_{\rm HF}(\mathbf{r})= {\rm Det}\left[e^{\mathbf{i}\mathbf{k}_i\cdot \mathbf{r}_j} 
    u_{\mathbf{k}_i}(\mathbf{r}_j)\right] .
\end{gathered}
\label{eq:hf_wf}
\end{equation}
Electrons are treated independently of each other with a mean-field interaction in Eq.~\eqref{eq:hf_wf}, simplifying quantum many-body problems but also deviating from the ground truth. 
To fully treat the electron correlation effects, we employ a correlated neural network wavefunction $\psi_{\rm net}$ from DeepSolid \cite{deepsolid}, whose general form reads
\begin{equation}
\begin{gathered}
    \psi_{\rm net}(\mathbf{r})={\rm Det}\left[e^{\mathbf{i}\mathbf{k}_i\cdot \mathbf{r}_j} 
    u_{\mathbf{k}_i}(\mathbf{r}_j;\mathbf{r}_{\neq j})\right],
\end{gathered}
\label{eq:wf}
\end{equation}
where $\mathbf{r}_{\neq j}$ denotes all the electron coordinates except $\mathbf{r}_j$. Eq.~\eqref{eq:wf} resembles the form of the traditional Bloch function, while cell-periodic functions $u_\mathbf{k}$ are now represented using deep neural networks that rely on all electrons to accommodate electron correlations \cite{ferminet}. 
Electron features $\mathbf{r}_i$ are converted to be periodic and permutation equivariant before being fed into neural networks, and complex-valued orbitals $u_\mathbf{k}$ are constructed with a pair of neural networks outputting the real and imaginary part respectively. As a result, Fermionic anti-symmetry, periodicity, and complex-valued nature are all encoded in our network, promoting it to be a legitimate and expressive ansatz for solid. See Ref.~\cite{deepsolid} for more details of the architecture.

Using the neural network we have constructed, the enthalpy functional outlined in Eq.~\eqref{eq:loss} can be efficiently minimized through variational Monte Carlo, allowing for gradual convergence to the ground truth. However, there have been significant fluctuations observed in $\hat{U}_i$ evaluation, which seriously impedes optimization. As a solution, antithetic sampling is employed in the Monte Carlo evaluation, which reads
\begin{equation}
\begin{gathered}
    \langle \hat{U}_i\rangle=
    \frac{\int d\mathbf{r}\ |\psi(\mathbf{r})|^2\ U_i(\mathbf{r})}{\int d\mathbf{r}\ |\psi(\mathbf{r})|^2} = \frac{\int d\mathbf{r}\ |\psi(\mathbf{r})|^2\ \widetilde{U}_i(\mathbf{r})}{\int\ d\mathbf{r}\ |\psi(\mathbf{r})|^2}, \\
    \widetilde{U}_i(\mathbf{r})=\frac{1}{2}\left[U_i(\mathbf{r})+\frac{|\psi(-\mathbf{\mathbf{r}})|^2}{|\psi(\mathbf{r})|^2}U_i(-\mathbf{r})\right].
\end{gathered}
\label{eq:anti}
\end{equation}
And thus the fluctuations are significantly reduced through the cancellation between $U_i(\mathbf{r})$ and its inverted image $U_i(-\mathbf{r})$.
It's worth noting that centrosymmetric cells are assumed in Eq.~\eqref{eq:anti}, and one can choose other images for cancellation if central symmetry is not satisfied.
To further improve efficiency, we have employed a Kronecker-factored curvature estimator (KFAC) optimizer \cite{kfac}, which effectively integrates second-order information into the optimization process, surpassing traditional optimizers. 
%Presented below are our results on a range of systems, including isolated atoms, one-dimensional chains, two-dimensional slabs, and three-dimensional alkali metal hydrides. Large deviations are found in DFT for these systems, and more accurate quantum chemistry methods are usually inapplicable, making them ideal trial scenarios for neural networks. 
See the Supplementary Material for more computational details, and the code of this work is developed at the open-source repository of DeepSolid \footnote{\url{https://github.com/bytedance/DeepSolid}}.

Isolated atoms are the first systems selected for direct comparison with the most accurate methods and experimental data. 
%Straightforward as the systems seem for the traditional methods, it should be noted that Coulomb interaction exhibits long-range behavior, necessitating the incorporation of diffusive basis functions for accurate polarization calculations. 
In our calculations, we place a single atom in a large enough box to eliminate periodic image interactions. The calculated polarizability $\alpha$ is shown in Tab.~\ref{tab:atom}, which measures the linear response of the dipole moment to the applied field and has some subtle relation with the bulk susceptibility $\chi$ (see Supplementary Material). Results from DFT with the B3LYP functional, HF, and CCSD(T) under OBC are also listed for comparison.
P-state atoms (B, C, O, F) are skipped because their anisotropy requires special treatments \cite{cc_second_period}.
As can be seen from Tab.~\ref{tab:atom}, although B3LYP is widely-trusted functional belonging to the fourth rung of the so-called Jacob's ladder of DFT, it consistently deviates from the ground truth and has a relatively large mean absolute error (MAE).
The behavior of DFT is due to the inaccuracy in treating the exchange-correlation effects, which can be very different for energy and polarization calculations.
In HF calculations, because of the explicit treatment of non-local exchange, deviations in polarization are significantly reduced. 
CCSD(T) is the coupled cluster theory with single, double, and perturbative triple excitations, and is considered a very accurate method in the literature. It further incorporates correlation effects on top of HF wavefunction and achieves smaller MAE than HF results.
Overall, DeepSolid results are comparable with CCSD(T), showing that the exchange-correlation treatments in our neural network are accurate and reliable for polarization calculations. 
\begin{figure}[b]
\centering
\includegraphics[width=0.5\textwidth]{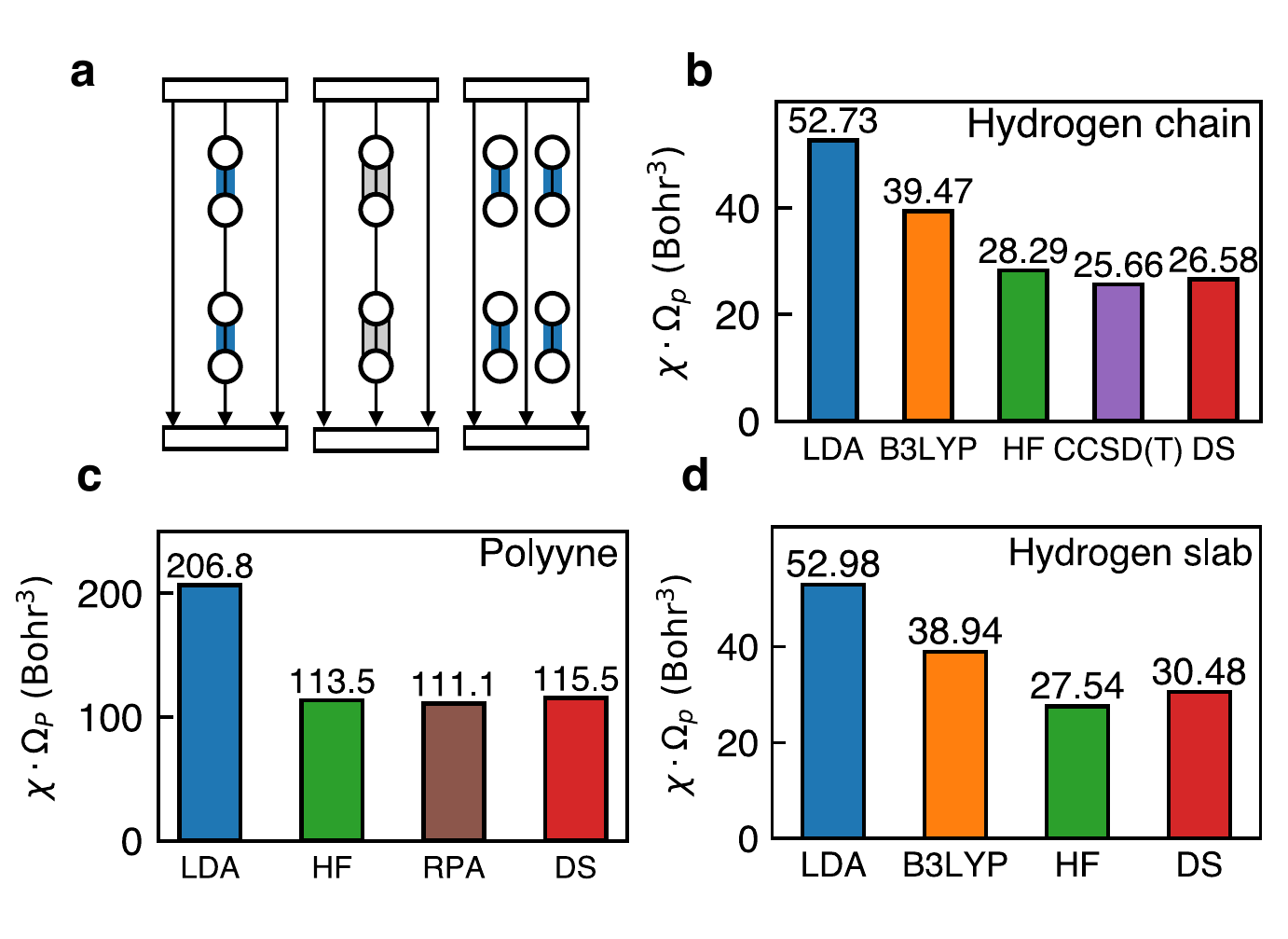}
\caption{\label{fig:hydrogen} Calculations of chains and slabs. DeepSolid results are labeled as DS.  $\mathbf{(a)}$ illustrations of hydrogen chain, polyyne, hydrogen slab, and the applied electric field. $\mathbf{(b)}$ hydrogen chain, $\mathbf{(c)}$ polyyne, $\mathbf{(d)}$ hydrogen slab susceptibilities $\chi$. $\Omega_p$ denotes the volume of the primitive cell. For the hydrogen system, LDA and HF results were calculated with the 3-21G basis set under PBC \cite{chain_slab_dft}. CCSD(T) calculations were performed with the 6-311G** basis set under OBC \cite{chain_qc}. Intrapair and interpair distances are set to 2 and 3 Bohr respectively for the hydrogen chain. The interchain distance is set to 4.724 Bohr for the slab.
For polyyne, the alternating distance between carbon atoms is set to 1.18 and 1.4 $\AA$. LDA, HF, and RPA results were calculated with 6-31G, 3-21G, and 4-31G basis set respectively under OBC \cite{polyyne_lda,polyyne_HF,polyyne_rpa}.}
\end{figure}

Having demonstrated our technique with single atoms, we proceed to simulate periodic systems by arranging bonded molecules into a one-dimensional chain and a two-dimensional slab. These systems are widely known as challenging cases for conventional DFT methods, which would have a serious overestimation of their longitudinal susceptibility. This problem stems from the fact that surface charges are insensitive to the bulk charge within the system when non-local interactions are absent, and this can be solved using more accurate \textit{ab initio} methods \cite{chain_slab_dft}. For the one-dimensional case, hydrogen chain (${\rm n\ H_2}$) and polyyne (${\rm n\ C_2}$) are studied, and the simulation size is pushed to ${\rm 22\ H_2}$ and ${\rm 7\ C_2}$ respectively for TDL convergence. Correlation-consistent effective core potential (ccECP) is employed for polyyne to accelerate neural network optimization and reduce fluctuation \cite{ccECP,ferminet_ecp}. The final results are plotted in Fig.~\ref{fig:hydrogen}, which show that susceptibility calculated by DeepSolid agrees well with correlated wavefunction methods CCSD(T) and random phase approximation (RPA).  
Local-density approximation (LDA) functional deviates severely from the ground truth for one-dimensional chains \cite{chain_qc}, but the use of hybrid functions such as B3LYP leads to partial recovery of non-local exchange effects and, consequently, a reduction in the overshot.
HF is much better than DFT calculations, which further proves the importance of the non-local exchange effect for electric polarization calculation in this system. 
As we arrange hydrogen chains periodically to form hydrogen slabs, the computational cost of high-level deterministic wavefunction methods, such as CCSD(T), grows rapidly and is soon beyond reach.
However, our approach has a lower scaling and we can obtain the first accurate polarization calculation for such a hydrogen slab (Fig.~\ref{fig:hydrogen}d).
And for the slab, the performances of DFT and HF compared with our accurate neural network results are similar to those observed for the chains. 

\begin{figure}[t]
\centering
\includegraphics[width=0.45\textwidth]{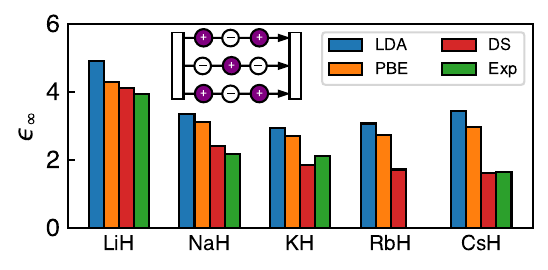}
\caption{\label{fig:XH} Calculated high-frequency dielectric constant $\epsilon_\infty$ of alkali metal hydrides XH. DeepSolid results are labeled as DS. LDA and PBE results in PBC are taken from Ref.~\cite{xh_dft}. Experimental data are taken from Refs.~\cite{lih_exp, csh_exp, xh_exp}, which are derived from refractive indexes $n_{\text{D}}$ for sodium doublet ($589.29\ {\rm nm}$) via Eq.~\eqref{eq:dielectric}. RbH experiments are absent.}
\end{figure}

% In addition to modeling hydrogen systems, our simulation is expanded to include carbon-containing polyyne and graphene, which possess a more intricate electronic structure and are of experimental interest. The longitudinal susceptibility of polyyne has been the focus of research for several decades \cite{polyyne_HF,polyyne_lda,polyyne_rpa,polyyne_ccsd}, however, a definitive conclusion has yet to be reached. To address this, we increased our simulation size to ${\rm 7\ C_2}$ chains and utilized a correlation-consistent effective core potential\cite{ccecp}. The results are shown in Fig.~\ref{fig:hydrogen}c, indicating that our neural network outcomes match well with the random-phase approximation (RPA) and show superior performance compared to LDA, similar to the hydrogen case. As for the two-dimensional graphene, there is significant interest in its dielectric response, which is closely related to superconductivity. In our simulation, we calculated the out-of-plane polarizability $\alpha_\perp$ of a $2\times 2$ supercell with twist average technique to minimize finite-size errors \cite{tabc}. The results are presented in Fig.~\ref{fig:hydrogen}d, where LDA and DMC results have also been included for comparison purposes \cite{graphene_alpha}. In contrast to previous longitudinal results, LDA agrees well with both DMC and neural network results, manifesting the independent response between electrons to the perpendicular electric field.

To further test our method, we applied it to alkali metal hydrides and calculated their dielectric constants, allowing direct comparison with experimental results. These systems have a simple structure, consisting of alternating cations and anions, but they are of considerable research significance due to their relevance in hydrogen storage applications \cite{hydrogen_storage}. %Recalling classical electromagnetism, 
The high-frequency dielectric constant $\epsilon_\infty$ can be extracted through optical experiments from the following relations,
\begin{equation}
\begin{gathered}
    \mathbf{D}=\epsilon_\infty \mathbf{E}=\mathbf{E}+4\pi\mathbf{P}\ , \\
    \epsilon_\infty = 1 + 4\pi \chi = n_{\text{D}}^2\ ,
\end{gathered}
\label{eq:dielectric}
\end{equation}
where $n_{\text{D}}$ denotes the corresponding refractive index. In the visible light regime, ions are almost frozen relative to the incident light frequency and this leads to the dominance of electric polarization in $\epsilon_\infty$.
These three-dimensional systems are also qualitatively different from the chains and slabs, because the fluctuation of $\hat{U}_i$ increases as we tile the cells in all three direction to form the simulation cell.
%During our simulation, large fluctuations of $\hat{U}_i$ are observed when the simulation cell is tiled in all three directions. 
To balance the influence from finite-size error and $\hat{U}_i$ fluctuations, we tile the conventional cell in the direction of the applied field $\mathbf{E}$ and the transverse directions remain unchanged. Moreover, Burkatzki-Filippi-Dolg (BFD) pseudopotential \cite{bfd_pp} is used to remove inertial core electrons \cite{ferminet_ecp}. 
Our calculations have been pushed to the $4\times1\times1$ supercell and the results are plotted in Fig.~\ref{fig:XH}. LDA and Perdew–Burke-Ernzerhof (PBE) results are also plotted for comparison, while more accurate conventional wavefunction methods are not applicable due to computational costs. 
As we can see, numerical simulations and experiments agree that $\epsilon_\infty$ decreases as the alkali metal atom becomes heavy, since $\epsilon_\infty$ is inversely proportional to the cell volume in Eq.~\eqref{eq:dielectric}. However, LDA and PBE functionals \cite{pbe} tend to overestimate $\epsilon_\infty$, and the error is largest in CsH. 
In contrast, our DeepSolid results agree well with the experiment for all systems, which manifests the capability of neural network wavefunction to capture non-local exchange and correlation effects.

\begin{figure}[b]
\centering
\includegraphics[width=0.5\textwidth]{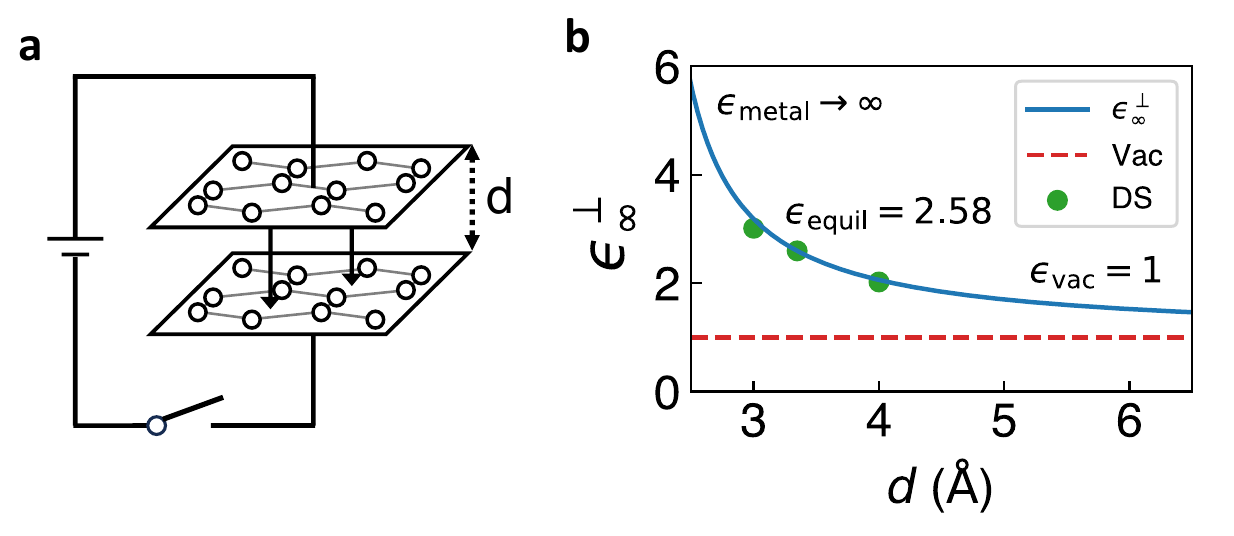}
\caption{\label{fig:2d_dielectric} Calculated effective 2D dielectric constant $\epsilon^\perp_\infty$ of bilayer graphene. ($\mathbf{a}$) plot of bilayer graphene under electric field. ($\mathbf{b}$) calculated $\epsilon^\perp_\infty$ as a function of graphene layer distance $d$. The equilibrium separation of BLG is set to 3.347 $\AA$. }
\end{figure}

After demonstrating the accuracy of our methods in previous sections, we now proceed to apply our method to bilayer graphene (BLG), an extensively studied 2D material system known for its rich electronic properties. Despite its fundamental importance, the precise value of the dielectric constant of BLG remains elusive and has been an important subject of both experimental and theoretical works  \cite{graphene_prb,graphene_alpha}. 
Specifically, theoretical calculations reported were either restricted to DFT level \cite{graphene_prb} or based on values calculated with monolayer graphene \cite{graphene_alpha}. Here we use DeepSolid to directly calculate the out-of-plane dielectric constant $\epsilon_\infty^{\perp}$ of bilayer graphene. 
%and re-establish $\epsilon_\infty^{\perp}$ changes as we tune the BLG layer distance $d$. 
%Specifically, 
$2\times2$ supercells containing monolayer and equilibrium AA-stacked bilayer graphene were used. 
The calculated monolayer polarizability equals $5.7(1)\ {\rm Bohr^3}$ and bilayer polarizability equals $11.6(1)\ {\rm Bohr^3}$, which agrees with the linear dependence of polarizability on the number of layers as shown in Ref.~\cite{graphene_prb}. 
Based on this linear dependence and following Ref.~\cite{bn_npj}, one can derive the expression of the out-of-plane dielectric constant as a function of the layer separation $d$ (see Supplementary Material):
\begin{equation}
\begin{gathered}
    \epsilon_\infty^{\perp}(d)=\left(1-\frac{2\pi\alpha^{\rm BLG}_{\rm equil}}{S\cdot d}\right)^{-1}\ ,
\end{gathered}
\label{eq:2d_dielectric}
\end{equation}
where $S=5.25\ \AA^2$ denotes the area of the primitive cell. 
Using the computed polarizability $\alpha^{\rm BLG}_{\rm equil}$, we can re-establish the relation of $\epsilon_\infty^{\perp}$, which is plotted in Fig.~\ref{fig:2d_dielectric}. To further check Eq.~\eqref{eq:2d_dielectric}, we also calculate the polarizability of bilayer graphene at slightly larger ($4\AA$) and smaller ($3\AA$) layer distances and plot the corresponding dielectric constant in Fig.~\ref{fig:2d_dielectric}, and the results agrees well with each other.
Moreover, there are two notable limits when varying the layer separation $d$: as $d$ decreases, two graphene layers coincide with each other and the system becomes metallic, which explains the diverging of $\epsilon_\infty^\perp$; as $d$ becomes large, BLG polarization becomes negligible and vacuum contribution dominants in $\epsilon_\infty^\perp$ which approaches unity.
The thickness-dependent dielectric constant will be valuable for further understanding and tuning the stacked multilayer graphene systems.

In conclusion, this work proposes an efficient and accurate method for investigating solid polarization based on the recently developed solid neural network wavefunction combined with quantum Monte Carlo. Our approach demonstrates superiority over the conventional state-of-the-art electronic structure methods. In the future, with the proposed framework, it is promising to investigate a wide range of phenomena, including ferroelectricity, topological electronic transport, quantum Hall effect, and orbital magnetization, among others, on a higher level of accuracy and with electron correlations accounted for properly. Furthermore, this work provides more possibilities for utilizing neural network applications in condensed matter physics.

\section*{Acknowledgements}

We want to thank ByteDance Research Group for inspiration and encouragement. This work is directed and supported by Hang Li and ByteDance Research. J.C. is supported by the National Natural Science Foundation of China under Grant No. 92165101.

\bibliography{reference}% Produces the bibliography via BibTeX.

\end{document}

% \section*{Appendix}\label{app}
% \title{Supplementary information: Macroscopic electric polarization from many-body neural network wavefunction}

\onecolumngrid
\section*{Supplementary Material}

\section{Gradient Formula}
DeepSolid ansatz can be straightforwardly optimized via enthalpy minimization, and the gradient reads
\begin{equation}
\begin{gathered}
    \nabla_\theta \langle F\rangle=\mathbf{G}_1+\mathbf{G}_2\ , \\
    \mathbf{G}_1={\rm Re}\left[\langle \hat{H}_S\nabla_\theta\ln\psi^*\rangle - \langle \hat{H}_S\rangle\langle\nabla_\theta\ln\psi^*\rangle\right]\ , \\
    \mathbf{G}_2=\mathbf{E}\cdot\sum_i\frac{\mathbf{a}_i}{2\pi}{\rm Re}\left[\left\langle{\rm Im}\left[\frac{\hat{U}_i}{\langle \hat{U}_i\rangle}\right]\nabla_\theta\ln\psi^*\right\rangle\right]\ .
\end{gathered}
\label{eq:grad}
\end{equation}
Note that antithetic sampling is used in $\langle\hat{U}_i\rangle$ evaluation.

\section{Quantity Definition}
The electric susceptibility of materials is defined as
\begin{equation}
    \chi=\frac{\partial \mathbf{P}}{\partial \mathbf{E}}\bigg\rvert_{\mathbf{E}=0}\ ,
\end{equation}
where $\mathbf{E}$ is the macroscopic electric field in the medium. Polarizability for isolated atoms and molecules is defined as
\begin{equation}
    \alpha=\frac{\partial \mathbf{p}}{\partial \mathbf{E}_{\rm ext}}\bigg\rvert_{\mathbf{E}_{\rm ext}=0}\ ,
\end{equation}
where ${\bf p}$ is the induced dipole moment and ${\bf E}_\text{ext}$ denotes the external electric field.
Despite the difference, the enthalpy formula is still applicable for isolated atoms, simply by replacing $\Omega_S\mathbf{P}$ with ${\bf p}$ and ${\bf E}$ with ${\bf E}_\text{ext}$. 

It's worth noting that susceptibility or polarizability receives non-linear contribution under finite electric fields, which reads
\begin{equation}
\label{eq:gamma}
    \frac{\partial \mathbf{p}}{\partial \mathbf{E_{\rm ext}}}=\alpha+\frac{1}{6}\gamma\mathbf{E}_{\rm ext}^2,
\end{equation}
where $\gamma$ denotes the hyperpolarizability of the system and higher order term are omitted. 
The applied electric fields should be small enough to neglect non-linear contribution.

Isolated slab polarizability is related to susceptibility $\chi$ as follows \cite{graphene_prb}
\begin{equation}
    \alpha_{\rm slab}=\Omega_p\frac{\chi}{1+4\pi\chi}=\frac{\Omega_p}{4\pi}\left(1-\frac{1}{\epsilon^\perp_{\rm SC}}\right) ,
\end{equation}
where $\epsilon^\perp_{\rm SC}$ denotes the dielectric constant of supercell perpendicular to the slab. $\epsilon^\perp_{\rm SC}$ received contributions from vacuum and contained 2D slab, which can be formulated as below \cite{bn_npj}
\begin{equation}
    \begin{gathered}
        \frac{c}{\epsilon^\perp_{\rm SC}} = \frac{c-t}{\epsilon_{\rm vac}}+\frac{t}{\epsilon_{\rm 2D}^\perp},
    \end{gathered}
\end{equation}
where $c$ denotes the perpendicular length of supercell and $t$ denotes the thickness of 2D material. The effective dielectric constant $\epsilon_{\rm 2D}^\perp$ of 2D material can then be derived
\begin{equation}
    \begin{gathered}
        \epsilon_{\rm 2D}^\perp = \left[1+\frac{c}{t}\left(\frac{1}{\epsilon_{\rm SC}}-1\right)\right]^{-1}=\left(1-\frac{4\pi\alpha_{\rm slab}}{A\cdot t}\right)^{-1}\ ,
    \end{gathered}
\end{equation}
where $A$ denotes the unit area of primitive cell. For bilayer material, $t=2d$ and $d$ denotes the layer distance.

\section{Hyperparameters}
The specific architecture of the neural network can be seen in Ref.~\cite{deepsolid}, and hyperparameters used in simulations are listed in Tab.~\ref{tab:ge_hyper}. The main difference from previous studies is the enlarged learning rate decay, which affects the learning rate as follows
\begin{equation}
\begin{gathered}
    {\rm lr}_i = \frac{{\rm lr}_0}{(1+i/{\rm delay})^{{\rm decay}}},
\end{gathered}
\label{eq:lr_decay}
\end{equation}
where lr denotes the learning rate, and $i$ counts the training steps. Learning rate decay is enlarged to stabilize the train since polarizability $\alpha$ and susceptibility $\chi$ appear sensitive to fluctuations, smaller lr has the nearly same results with larger fluctuations in training. Part of the hyperparameters are different for test systems and will be clarified respectively. Moreover, Float64 is necessary for antithetic sampling.
\begin{table}[t]
\centering
\caption{Recommended hyperparameters
}
\begin{ruledtabular}
\begin{tabular}{cccc}
Hyperparameter & Value & Hyperparameter & Value\\
\hline
Pretrain basis & NA & Pretrain iterations  & NA\\  % chktex 8
Dimension of one-electron layer $\mathbf{V}$ & 256 & Dimension of two-electron layer $\mathbf{W}$ & 32 \\
Number of layers  & 4 & Number of determinants & 8\\
Optimizer & KFAC & Learning rate & 1e-2\\
Learning rate decay & 2 & Learning rate delay & 1e4 \\
Damping & 1e-3 & Constrained norm of gradient & 1e-3 \\
Momentum of optimizer & 0.0 & Batch size & 4096 \\
Number of training steps & 1e5 & Clipping window of gradient & 5 \\
MCMC burn in & 1e3 & MCMC steps between each iteration & 30 \\
MCMC move width & 2e-2 & Target MCMC acceptance & 55\% \\
Precision & Float64 & Number of inference steps & 2e4 \\  
\end{tabular}
\end{ruledtabular}
\label{tab:ge_hyper}
\end{table}

\section{Atoms}
\begin{figure}[H]
\centering
\includegraphics[width=1\textwidth]{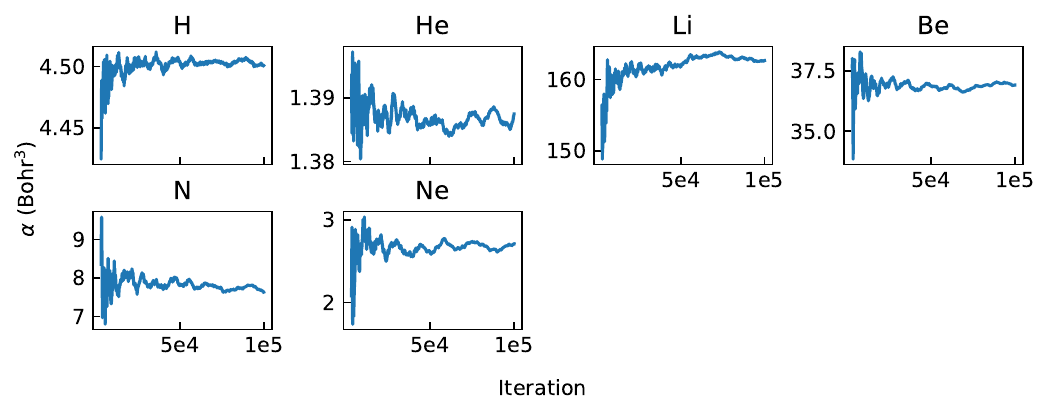}
\caption{\label{fig:atom_train} Training curves of atom polarizabilities. For clarity, at each iteration number, we plot the median value of the last 10$\%$ of the corresponding iteration.}
\end{figure}

Single atoms are placed in a cubic cell with a length of 2000 Bohr to neglect image interactions. The applied electric fields are given in Tab.~\ref{tab:E_atom}. According the the hyperpolarizability calculated in Refs.~\cite{h_gamma,he_gamma,li_gamma, be_gamma, hyperpolar} and Eq.~\eqref{eq:gamma}, non-linear contributions to polarizability $\alpha$ caused by applied electric field are less than 0.01 ${\rm Bohr^3}$. Moreover, a larger learning rate (1e-1) and gradient clip window (10) are necessary to reach the energy minimum of lithium and beryllium. The training curves of polarizability are given in Fig.~\ref{fig:atom_train}.
\begin{table}[H]
\caption{\label{tab:E_atom} Applied electric field $\mathbf{E}$ for atoms in atomic unit.}
\begin{ruledtabular}
\begin{tabular}{ccccccccccc} 
    %\hline
      & H & He & Li & Be & N & Ne\\
     \hline
    $\mathbf{E}$ & 5e-3 & 1e-2 & 1e-3 & 2e-3 & 5e-3 & 1e-2 
\end{tabular}
\end{ruledtabular}
\end{table}

\section{Hydrogen Chain}
\begin{figure}[H]
\centering
\includegraphics[width=1\textwidth]{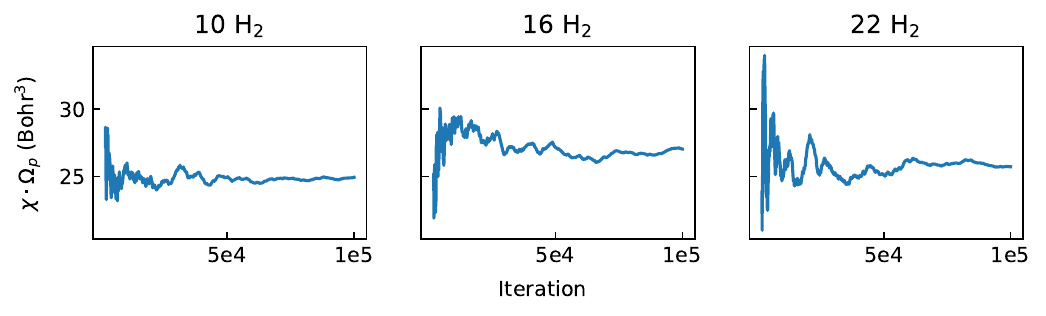}
\caption{\label{fig:hydrogen_chain_train} Training curves of hydrogen chain susceptibilities. For clarity, at each iteration number, we plot the median value of the last 10$\%$ of the corresponding iteration.}
\end{figure}

Hydrogen chain geometry is set to be centrosymmetric for antithetic sampling, and it is given in Tab.~\ref{tab:hydrogen_chain_geo}. The applied field is set to 1e-3 atomic unit. According the the hypersusceptibility calculated in Ref.~\cite{chain_slab_dft}, non-linear contributions to susceptibility caused by finite electric field are less than 0.01 ${\rm Bohr^3}$. Simulation size ranges from ${\rm 10\ H_2}$ to ${\rm 22\ H_2}$ to ensure TDL convergence and corresponding susceptibility $\chi$ is given in Tab.~\ref{tab:hydrogen_chain_alpha}. The training curves of susceptibility are given in Fig.~\ref{fig:hydrogen_chain_train}.
\begin{table}[H]
\caption{\label{tab:hydrogen_chain_geo} Geometry of hydrogen chain.}
\begin{ruledtabular}
\begin{tabular}{cccc} 
    %\hline
      Atom & Position (Bohr) & Lattice vector & Value (Bohr) \\
      \hline
      H1 & (1.5, 0, 0) & $\mathbf{a}_1$ & (5, 0, 0) \\
      H2 & (3.5, 0, 0) & $\mathbf{a}_2$ & (0, 100, 0) \\
      & & $\mathbf{a}_3$ & (0, 0, 100) \\
\end{tabular}
\end{ruledtabular}
\end{table}
\begin{table}[H]
\caption{\label{tab:hydrogen_chain_alpha} Longitudinal linear susceptibility of hydrogen chain.}
\begin{ruledtabular}
\begin{tabular}{cccc} 
    %\hline
      System & ${\rm 10\ H_2}$ & ${\rm 16\ H_2}$ & ${\rm 22\ H_2}$ \\
      \hline
      $\chi\cdot\Omega_p$ $({\rm Bohr}^3)$ & 25.21(1) & 27.37(2) & 26.58(2) \\
\end{tabular}
\end{ruledtabular}
\end{table}

\section{Hydrogen Slab}
\begin{figure}[H]
\centering
\includegraphics[width=1\textwidth]{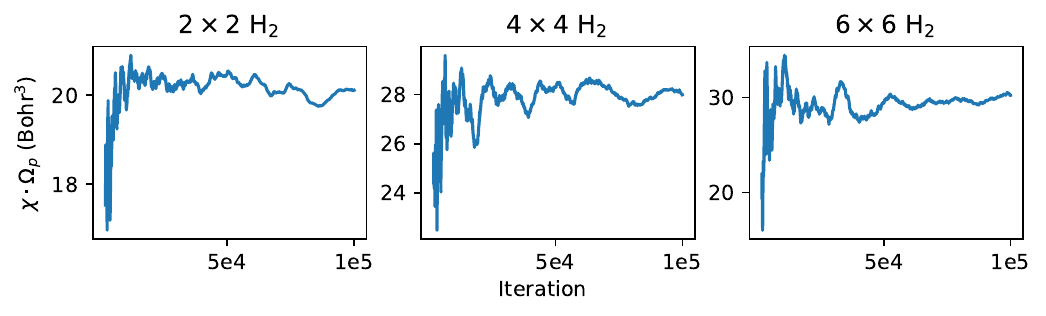}
\caption{\label{fig:hydrogen_slab_train} Training curves of hydrogen slab susceptibilities. For clarity, at each iteration number, we plot the median value of the last 10$\%$ of the corresponding iteration.}
\end{figure}

The specific geometry of the hydrogen slab is given in Tab.~\ref{tab:hydrogen_slab_geo}. The applied field is set to 2e-3 atomic unit.  According the the hypersusceptibility calculated in Ref.~\cite{chain_slab_dft}, non-linear contributions contributions caused by applied electric field are less than 0.04 ${\rm Bohr^3}$. Simulation size ranges from ${\rm 2\times 2 \ H_2}$ to ${\rm 6\times 6\ H_2}$ to ensure TDL convergence and corresponding susceptibility $\chi$ is given in Tab.~\ref{tab:hydrogen_slab_alpha}. The training curves of susceptibility is given in Fig.~\ref{fig:hydrogen_slab_train}.
\begin{table}[H]
\caption{\label{tab:hydrogen_slab_geo} Geometry of hydrogen slab.}
\begin{ruledtabular}
\begin{tabular}{cccc} 
    %\hline
      Atom & Position (Bohr) & Lattice vector & Value (Bohr) \\
      \hline
      H1 & (1.5, 0, 0) & $\mathbf{a}_1$ & (5, 0, 0) \\
      H2 & (3.5, 0, 0) & $\mathbf{a}_2$ & (0, 4.724, 0) \\
      & & $\mathbf{a}_3$ & (0, 0, 100) \\
\end{tabular}
\end{ruledtabular}
\end{table}
\begin{table}[H]
\caption{\label{tab:hydrogen_slab_alpha} Susceptibility of hydrogen slab.}
\begin{ruledtabular}
\begin{tabular}{cccc} 
    %\hline
      System & ${\rm 2\times2 \ H_2}$ & ${\rm 4\times 4\ H_2}$ & ${\rm 6\times 6\ H_2}$ \\
      \hline
      $\chi\cdot\Omega_p$ $({\rm Bohr}^3)$ & 20.44(1) & 28.76(8) & 30.5(2) \\
\end{tabular}
\end{ruledtabular}
\end{table}

\section{Polyyne}
\begin{figure}[H]
\centering
\includegraphics[width=0.8\textwidth]{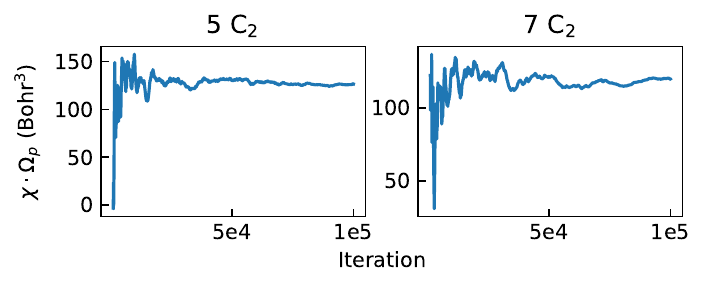}
\caption{\label{fig:polyyne_train} Training curves of polyyne susceptibilities. For clarity, at each iteration number, we plot the median value of the last 10$\%$ of the corresponding iteration.}
\end{figure}

Polyyne ${\rm n\ C_{2}}$ geometry is given in Tab.~\ref{tab:poly_geo}. The applied field is set to 1e-3 atomic unit. According the the hypersusceptibility calculated in Ref.~\cite{polyyne_HF}, non-linear contributions contributions caused by applied electric field are less than 0.2 ${\rm Bohr^3}$. Simulation size is pushed to ${\rm n=7}$ and corresponding susceptibility $\chi$ is given in Tab.~\ref{tab:hydrogen_chain_alpha}. The training curves of susceptibility are given in Fig.~\ref{fig:polyyne_train}.
\begin{table}[H]
\caption{\label{tab:poly_geo} Geometry of polyyne.}
\begin{ruledtabular}
\begin{tabular}{cccc} 
    %\hline
      Atom & Position ($\AA$) & Lattice vector & Value ($\AA$) \\
      \hline
      C1 & (0.7, 0, 0) & $\mathbf{a}_1$ & (2.58, 0, 0) \\
      C2 & (1.88, 0, 0) & $\mathbf{a}_2$ & (0, 100, 0) \\
      & & $\mathbf{a}_3$ & (0, 0, 100) \\
\end{tabular}
\end{ruledtabular}
\end{table}

\begin{table}[H]
\caption{\label{tab:poly_alpha} Longitudinal linear susceptibility of polyyne.}
\begin{ruledtabular}
\begin{tabular}{ccc} 
    %\hline
      System & ${\rm 5\ C_{2}}$ & ${\rm 7\ C_{2}}$  \\
      \hline
      $\chi\cdot \Omega_p$ $({\rm Bohr}^3)$ & 125.6(6) & 115.5(3) \\
\end{tabular}
\end{ruledtabular}
\end{table}

\section{Alkali Metal Hydrides}
\begin{figure}[H]
\centering
\includegraphics[width=1\textwidth]{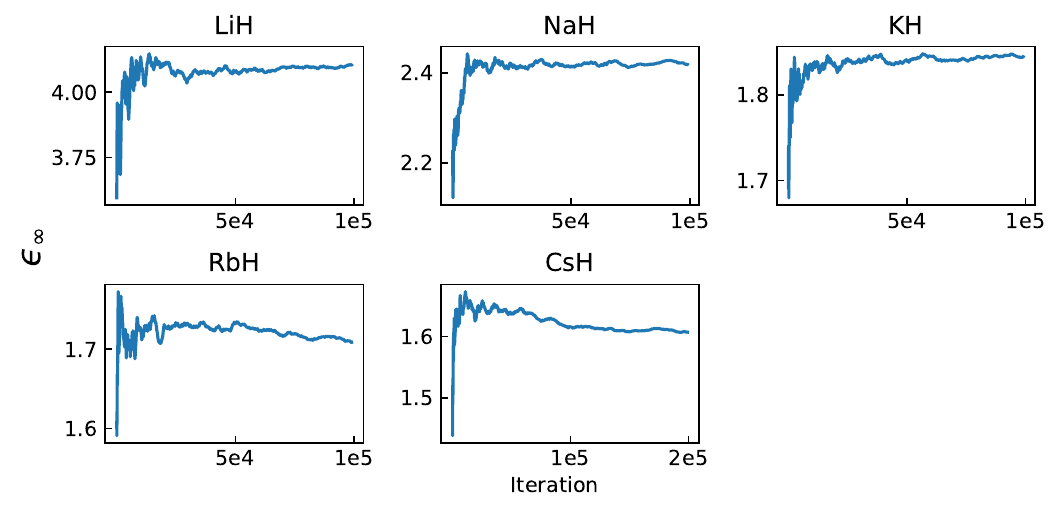}
\caption{\label{fig:XH_train} Training curves of alkali metal hydrides 
 dielectric constant $\epsilon_\infty$. For clarity, at each iteration number, we plot the median value of the last 10$\%$ of the corresponding iteration.}
\end{figure}

The specific geometry of alkali metal hydrides is given in Tab.~\ref{tab:XH_geo}. The lattice constant of each alkali metal is set to experiment data \cite{xh_dft} and is given in Tab.~\ref{tab:XH_length}, the corresponding applied electric field is also listed. Smaller electric field (2e-3 atomic unit) is also used to check finite field effect, and nonlinear contribution is less than 0.1 ${\rm Bohr^3}$ in susceptibility. High-frequency dielectric constants $\epsilon_\infty$ from DFT \cite{xh_dft}, DeepSolid(DS), and experiments \cite{lih_exp,csh_exp,xh_exp} are given in Tab.~\ref{tab:XH_dielectric}. The training curves of high-frequency dielectric constant $\epsilon_\infty$ are given in Fig.~\ref{fig:XH_train}. To ensure the TDL convergence, $\epsilon_\infty$ at different simulation sizes are given in Tab.~\ref{tab:XH_TDL}.
\begin{table}[H]
\caption{\label{tab:XH_geo} Geometry of alkali metal hydrides.}
\begin{ruledtabular}
\begin{tabular}{cccc} 
    %\hline
      Atom & Position & Lattice vector & Position \\
      \hline
      H1 & (L/2, 0, 0) & $\mathbf{a}_1$ & (L, 0, 0) \\
      H2 & (0, L/2, 0) & $\mathbf{a}_2$ & (0, L, 0) \\
      H3 & (0, 0, L/2) & $\mathbf{a}_3$ & (0, 0, L) \\
      H4 & (L/2, L/2, L/2) & & \\
      X1 & (0, 0, 0) & & \\
      X2 & (0, L/2, L/2) & & \\
      X3 & (L/2, 0, L/2) & & \\
      X4 & (L/2, L/2, 0) & & \\
\end{tabular}
\end{ruledtabular}
\end{table}
\begin{table}[H]
\caption{\label{tab:XH_length} Lattice constant and applied electric field $\mathbf{E}$ of alkali metal hydrides.}
\begin{ruledtabular}
\begin{tabular}{cccccc} 
    %\hline
      XH & LiH & NaH & KH & RbH & CsH \\
      \hline
      L($\AA$) & 4.086 & 4.89 & 5.704 & 6.037 & 6.388 \\
      $\mathbf{E}$ & 5e-3 & 5e-3 & 5e-3 & 5e-3 & 5e-3 \\
\end{tabular}
\end{ruledtabular}
\end{table}
\begin{table}[H]
\caption{\label{tab:XH_dielectric} High-frequency dielectric constant $\epsilon_\infty$ of alkali metal hydrides.}
\begin{ruledtabular}
\begin{tabular}{cccccc} 
    %\hline
      system & LiH & NaH & KH & RbH & CsH \\
      \hline
      LDA & 4.92 & 3.35 & 2.94 & 3.07 & 3.45 \\
      PBE & 4.28 & 3.12 & 2.69 & 2.73 & 2.98 \\
      DS & 4.128(3) & 2.422(1) & 1.8546(6) & 1.7162(4) & 1.607(1) \\
      Exp & 3.939 & 2.161 & 2.111 & NA & 1.638 
\end{tabular}
\end{ruledtabular}
\end{table}
\begin{table}[H]
\caption{\label{tab:XH_TDL} High-frequency dielectric constant $\epsilon_\infty$ of alkali metal hydrides at different size.}
\begin{ruledtabular}
\begin{tabular}{cccc} 
    %\hline
        $\epsilon_\infty$ & $2\times 1\times 1$ & $3\times 1\times 1$ & $4\times 1\times 1$\\
      \hline
      ${\rm LiH}$ & 3.913(3) & 4.059(2) & 4.128(3) \\
      ${\rm NaH}$ & 2.513(1) & 2.447(1) & 2.422(1) \\
      ${\rm KH}$ & 1.8951(5) & 1.8419(5) & 1.845(1) \\
      ${\rm RbH}$ & 1.7919(4) & 1.7315(4) & 1.7162(4) \\
      ${\rm CsH}$ & 1.6537(4) & 1.6156(4) & 1.607(1) 
\end{tabular}
\end{ruledtabular}
\end{table}

\section{Monolayer Graphene}
\begin{figure}[H]
\centering
\includegraphics[width=1\textwidth]{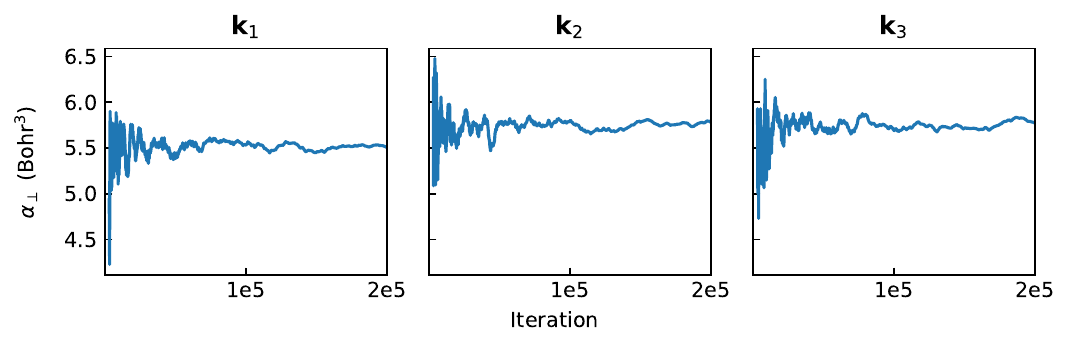}
\caption{\label{fig:graphene_train} Training curves of monolayer graphene slab polarizability. For clarity, at each iteration number, we plot the median value of the last 10$\%$ of the corresponding iteration.}
\end{figure}

Geometry of monolayer graphene is given in Tab~\ref{tab:graphene_geo}. Applied electric field is set to 5e-3 atomic unit. Smaller electric field (2e-3 atomic unit) is also used to check finite field effect, and nonlinear contribution is less than 0.1 ${\rm Bohr^3}$ in slab polarizability. Twist average is used to reduce finite size error and the formula reads
\begin{equation}
\label{eq:tabc}
    \begin{gathered}
    \alpha_{\rm avg}=\frac{1}{9}\alpha_{\mathbf{k}_1}+\frac{2}{3}\alpha_{\mathbf{k}_2}+\frac{2}{9}\alpha_{\mathbf{k}_3}\ , \\
    \mathbf{k_1}=0,\ \mathbf{k}_2=\frac{1}{3}\mathbf{b}_1+\frac{1}{3}\mathbf{b}_2,\ \mathbf{k}_3=\frac{2}{3}\mathbf{b}_1+\frac{1}{3}\mathbf{b_2}\ .
    \end{gathered}
\end{equation}
Corresponding training curve of polarizability $\alpha_\perp$ is given in Fig.~\ref{fig:graphene_train}. Final results is given in Tab.~\ref{tab:graphene_alpha}.
\begin{table}[H]
\caption{\label{tab:graphene_geo} Geometry of monolayer graphene.}
\begin{ruledtabular}
\begin{tabular}{cccc} 
    %\hline
      Atom & Position ($\AA$) & Lattice vector & Value ($\AA$) \\
      \hline
      C1 & (1.421, 0, 0) & $\mathbf{a}_1$ & (2.1315, -1.2306, 0) \\
      C2 & (2.842, 0, 0) & $\mathbf{a}_2$ & (2.1315, -2.1306, 0) \\
      & & $\mathbf{a}_3$ & (0, 0, 15.875) \\
\end{tabular}
\end{ruledtabular}
\end{table}

\begin{table}[H]
\caption{\label{tab:graphene_alpha} Slab poalrizability of graphene.}
\begin{ruledtabular}
\begin{tabular}{ccccc} 
    %\hline
        & $\mathbf{k_1}$ & $\mathbf{k_2}$ & $\mathbf{k_3}$ & Twist average\\
      \hline
      $\alpha_\perp({\rm Bohr}^3)$ & 5.5(1) & 5.7(1) & 5.7(1) & 5.7(1)
\end{tabular}
\end{ruledtabular}
\end{table}

\section{Bilayer Graphene}
\begin{figure}[H]
\centering
\includegraphics[width=1\textwidth]{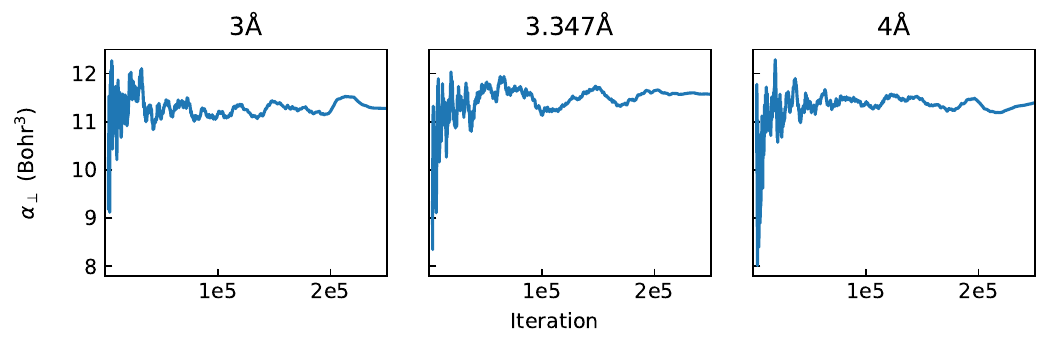}
\caption{\label{fig:bilayer_graphene_train} Training curves of bilayer graphene slab polarizability. For clarity, at each iteration number, we plot the median value of the last 10$\%$ of the corresponding iteration.}
\end{figure}

AA-stacked bilayer graphene is choosed for simulation and its geometry is given in Tab~\ref{tab:bilayer_graphene_geo}. And L is set to ${\rm 3.000\AA,\ 3.347\AA,\ and\ 4.000\AA}$ respectively. Applied electric field is set to 5e-3 atomic unit. Twist is set to $\mathbf{k}_S=\mathbf{b}_1/3+\mathbf{b}_2/3$ which has the largest weight in $3\times 3$ twist average. Moreover, the learning rate decay is set to 3 after 2e5 iterations to reduce fluctuations. Corresponding training curve of polarizability is given in Fig.~\ref{fig:bilayer_graphene_train}. The calculated polarizability is given in Tab.~\ref{tab:blg_alpha}. 

\begin{table}[H]
\caption{\label{tab:bilayer_graphene_geo} Geometry of bilayer graphene.}
\begin{ruledtabular}
\begin{tabular}{cccc} 
    %\hline
      Atom & Position ($\AA$) & Lattice vector & Value ($\AA$) \\
      \hline
      C1 & (1.421, 0, L/2) & $\mathbf{a}_1$ & (2.1315, -1.2306, 0) \\
      C2 & (2.842, 0, L/2) & $\mathbf{a}_2$ & (2.1315, -2.1306, 0) \\
      C3 & (1.421, 0, 30-L/2) & $\mathbf{a}_3$ & (0, 0, 15.875) \\
      C4 & (2.842, 0, 30-L/2) &  &  \\
\end{tabular}
\end{ruledtabular}
\end{table}

\begin{table}[H]
\caption{\label{tab:blg_alpha} Slab poalrizability of bilayer graphene.}
\begin{ruledtabular}
\begin{tabular}{cccc} 
    %\hline
        ${\rm L(\AA)}$ & ${\rm 3.000}$ & ${\rm 3.347}$ & ${\rm 4.000}$ \\
      \hline
      $\alpha_\perp({\rm Bohr}^3)$ & 11.3(1) & 11.6(1) & 11.4(1)
\end{tabular}
\end{ruledtabular}
\end{table}

\bibliography{supplement}